\documentstyle[12pt,epsfig]{article}
\begin{document}

\vspace{0.2cm}

\begin{center}
{\large\bf $D^0$-$\bar{D}^0$ Mixing and $CP$ Violation in $D^0$ vs
$\bar{D}^0 \rightarrow K^{*\pm} K^\mp$ Decays}
\end{center}

\vspace{.5cm}
\begin{center}
{\bf Zhi-zhong Xing} \footnote{E-mail: xingzz@mail.ihep.ac.cn} ~ and
~
{\bf Shun Zhou} \footnote{E-mail: zhoush@mail.ihep.ac.cn} \\
{\small\sl Institute of High Energy Physics, Chinese Academy of
Sciences, Beijing 100049, China}
\end{center}

\vspace{3.5cm}

\begin{abstract}
The noteworthy BaBar and Belle evidence for $D^0$-$\bar{D}^0$
mixing motivates us to study its impact on $D^0\rightarrow
K^{*\pm} K^\mp$ decays and their $CP$-conjugate processes. We show
that both the $D^0$-$\bar{D}^0$ mixing parameters ($x$ and $y$)
and the strong phase difference between $\bar{D}^0\rightarrow
K^{*\pm}K^\mp$ and $D^0\rightarrow K^{*\pm}K^\mp$ transitions
($\delta$) can be determined or constrained from the
time-dependent measurements of these decay modes. On the $\psi
(3770)$ and $\psi (4140)$ resonances at a $\tau$-charm factory, it
is even possible to determine or constrain $x$, $y$ and $\delta$
from the time-independent measurements of coherent $(D^0\bar{D}^0)
\rightarrow (K^{*\pm} K^\mp)(K^{*\pm} K^\mp)$ decays. If the
$CP$-violating phase of $D^0$-$\bar{D}^0$ mixing is significant in
a scenario beyond the standard model, it can also be extracted
from the $K^{*\pm} K^\mp$ events.
\end{abstract}

\vspace{0.4cm}

\begin{center}
PACS number(s): 11.30.Er, 12.15.Ff, 13.20.Fc, 13.25.Ft
\end{center}

\newpage

\framebox{\large\bf 1} ~ The BaBar \cite{BB} and Belle \cite{BE}
experiments have recently provided us with some noteworthy
evidence for $D^0$-$\bar{D}^0$ mixing, a quantum phenomenon
similar to $K^0$-$\bar{K}^0$, $B^0_d$-$\bar{B}^0_d$ or
$B^0_s$-$\bar{B}^0_s$ mixing. Both experiments indicate a
non-vanishing width difference between the mass eigenstates
$D^{}_1$ and $D^{}_2$ of $D^0$ and $\bar{D}^0$ mesons,
\begin{eqnarray}
y^\prime \cos\phi & = & \left (0.97 \pm 0.44 \pm 0.31 \right )
\times 10^{-2} \; , \nonumber \\
y^{}_{CP} & = & \left (1.31 \pm 0.32 \pm 0.25 \right ) \times
10^{-2} \; ,
\end{eqnarray}
where the values of $y^\prime \cos\phi$ and $y^{}_{CP}$ are
extracted from the decay modes $D^0 \rightarrow \bar{D}^0
\rightarrow K^+\pi^-$ versus $D^0 \rightarrow K^+\pi^-$ \cite{BB}
and $D^0\rightarrow \bar{D}^0 \rightarrow K^+K^-$ and $\pi^+\pi^-$
versus $D^0\rightarrow K^+K^-$ and $\pi^+\pi^-$ \cite{BE},
respectively. By linking $y^\prime$ and $y^{}_{CP}$ to the
$D^0$-$\bar{D}^0$ mixing parameters $x \equiv (M^{}_2 -
M^{}_1)/\Gamma$ and $y \equiv (\Gamma^{}_2 -
\Gamma^{}_1)/(2\Gamma)$ with $\Gamma = (\Gamma^{}_1 +
\Gamma^{}_2)/2$ and $\Gamma^{}_{1,2}$ being the width of
$D^{}_{1,2}$, Nir has pointed out that $|y| \sim 0.01$, $|x| <
|y|$ and small or vanishingly small $CP$ violation are expected in
the $D^0$-$\bar{D}^0$ mixing system within the standard model
\cite{Nir}. Some other authors have also discussed possible
implications of the BaBar and Belle measurements of
$D^0$-$\bar{D}^0$ mixing, either within or beyond the standard
model \cite{Italy}--\cite{Ball}.

Unfortunately, current theoretical calculations of
$D^0$-$\bar{D}^0$ mixing involve large uncertainties because of
the dominance of long-distance contributions \cite{LD}. In the
standard model, the values of $x$ and $y$ are expected to be a
second-order effect of the $SU(3)$ flavor symmetry breaking
\cite{Petrov}: $x,y \sim \sin^2\theta^{}_{\rm C} \times [SU(3) ~
{\rm breaking}]^2$, where $\theta^{}_{\rm C} \approx 13^\circ$
denotes the Cabibbo angle. A very reliable prediction for the size
of $SU(3)$ breaking has been lacking, although many attempts have
been made \cite{LD,Burdman}. Hence these two $D^0$-$\bar{D}^0$
mixing parameters might be only of limited use in testing the
standard model and searching for new physics. From an experimental
point of view, however, it is always desirable to measure or
constrain $x$ and $y$ as accurately as possible.

Motivated by the afore-mentioned positive results from the BaBar
and Belle experiments, here we aim to investigate the impact of
$D^0$-$\bar{D}^0$ mixing on $D^0 \rightarrow K^{*\pm} K^\mp$
decays and their $CP$-conjugate processes. Because $K^{*+}K^-$ (or
$K^{*-}K^+$) is not a $CP$ eigenstate, the amplitudes of $D^0$ and
$\bar{D}^0$ decays into $K^{*+}K^-$ (or $K^{*-}K^+$) may have a
significant strong phase difference $\delta$. In contrast, $D^0$
vs $\bar{D}^0\rightarrow K^+K^-$ decays do not involve such a
strong phase difference. We show that both the $D^0$-$\bar{D}^0$
mixing parameters ($x$ and $y$) and the strong phase difference
($\delta$) can be determined or constrained from the
time-dependent measurements of $D^0$ vs $\bar{D}^0 \rightarrow
K^{*\pm} K^\mp$ decays. On the $\psi (3770)$ and $\psi (4140)$
resonances at a $\tau$-charm factory (e.g., BEPC-II \cite{BEPC}),
we find that it is even possible to determine or constrain $x$,
$y$ and $\delta$ from the time-independent measurements of
coherent $(D^0\bar{D}^0) \rightarrow (K^{*\pm} K^\mp)(K^{*\pm}
K^\mp)$ events. If the $CP$-violating phase of $D^0$-$\bar{D}^0$
mixing is significant in a scenario beyond the standard model, it
can also be extracted from the decay modes under discussion.

The remaining part of this paper is organized as follows. Section
2 is devoted to the effects of $D^0$-$\bar{D}^0$ mixing and $CP$
violation in the time-dependent $D^0$ vs $\bar{D}^0 \rightarrow
K^{*\pm} K^\mp$ decays. The coherent $(D^0\bar{D}^0)\rightarrow
(K^{*\pm}K^{\mp})(K^{*\pm}K^{\mp})$ decays on the $\psi(3770)$ and
$\psi(4140)$ resonances are discussed in section 3, where we focus
our interest on possible signals of $D^0$-$\bar{D}^0$ mixing and
$CP$ violation in the time-independent events. An isospin analysis
of the final state interactions in $D\rightarrow KK^*$ modes is
done in section 4. Finally, we summarize our main results in
section 5.

\vspace{0.5cm}

\framebox{\large\bf 2} ~ In the standard model $D^0 \rightarrow
K^{*\pm} K^\mp$ transitions can occur through both tree-level and
loop-induced (penguin) quark diagrams. The former is essentially
$CP$-conserving (proportional to $V^{}_{cs}V^*_{us}$), while the
latter is negligibly small (suppressed by $m^2_q/M^2_W$ for $q=d,
s, b$) \cite{Nir}. Hence the four amplitudes $A^{}_{K^{*+}K^-}
\equiv \langle K^{*+}K^-|{\cal H}|D^0\rangle$,
$\bar{A}^{}_{K^{*+}K^-} \equiv \langle K^{*+}K^-|{\cal
H}|\bar{D}^0\rangle$, $A^{}_{K^{*-}K^+} \equiv \langle
K^{*-}K^+|{\cal H}|D^0\rangle$ and $\bar{A}^{}_{K^{*-}K^+} \equiv
\langle K^{*-}K^+|{\cal H}|\bar{D}^0\rangle$ have the relations
$\bar{A}^{}_{K^{*-}K^+} = A^{}_{K^{*+}K^-}$ and $A^{}_{K^{*-}K^+}
= \bar{A}^{}_{K^{*+}K^-}$ as a good approximation. We define
\begin{equation}
\frac{\bar{A}^{}_{K^{*+}K^-}}{A^{}_{K^{*+}K^-}} \; = \;
\frac{A^{}_{K^{*-}K^+}}{\bar{A}^{}_{K^{*-}K^+}} \; \equiv \; \rho
e^{i\delta} \; ,
\end{equation}
where $\rho >0$ and $\delta$ is the strong phase difference. On
the other hand, two neutral $D$-meson mass eigenstates can be
written as
\begin{eqnarray}
|D^{}_1\rangle & = & p|D^0\rangle + q|\bar{D}^0\rangle \; ,
\nonumber \\
|D^{}_2\rangle & = & p|D^0\rangle - q|\bar{D}^0\rangle \; ,
\end{eqnarray}
where $p$ and $q$ satisfy the normalization condition $|p|^2 +
|q|^2 = 1$. The phase of $q/p$ is $\phi \equiv \arg (q/p) = \arg
[(V^*_{cs}V^{}_{us})/(V^{}_{cs}V^*_{us})] \approx 0$ within the
standard model \cite{Xing96}, but it might be significant if a
kind of new physics contributes to the box diagram of
$D^0$-$\bar{D}^0$ mixing
\cite{Nir,Italy,Buras,He,Burdman,Grossman}. Allowing for both
$|q/p| \neq 1$ and $\phi \neq 0$, we may use the following
rephasing-invariant quantities to express the decay rates of $D^0
\rightarrow K^{*\pm} K^\mp$ and $\bar{D}^0 \rightarrow K^{*\pm}
K^\mp$:
\begin{eqnarray}
\lambda^{}_{K^{*+}K^-} & \equiv & \frac{q}{p} \cdot
\frac{\bar{A}^{}_{K^{*+}K^-}}{A^{}_{K^{*+}K^-}} \; = \; \rho \left
|\frac{q}{p} \right | e^{i\left
(\delta + \phi\right )} \; , \nonumber \\
\bar{\lambda}^{}_{K^{*-}K^+} & \equiv & \frac{p}{q} \cdot
\frac{A^{}_{K^{*-}K^+}}{\bar{A}^{}_{K^{*-}K^+}} \; = \; \rho \left
|\frac{p}{q} \right | e^{i\left (\delta - \phi\right )} \; .
\end{eqnarray}
Since the naive factorization approximation yields $\rho \sim
{\cal O}(1)$, it is quite natural to expect that
$|\lambda^{}_{K^{*+}K^-}| \approx |\bar{\lambda}^{}_{K^{*-}K^+}|
\sim {\cal O}(1)$ holds.

First, let us look at the time-dependent decay rates of $D^0$ vs
$\bar{D}^0 \rightarrow K^{*\pm}K^\mp$. Now that the
$D^0$-$\bar{D}^0$ mixing parameters $x$ and $y$ are both small, we
may just keep the terms of ${\cal O}(x)$ and ${\cal O}(y)$ in our
calculations. Using the generic formulas given in Ref.
\cite{Xing97}
\footnote{Note that $y = -y^{}_D$, where $y^{}_D$ is the
$D^0$-$\bar{D}^0$ mixing parameter defined in Ref.
\cite{Xing97}.},
we explicitly obtain
\begin{eqnarray}
\Gamma [D^0(t) \rightarrow K^{*+}K^-] & \propto &
|A^{}_{K^{*+}K^-}|^2 e^{-\Gamma t} [ 1 + (y {\rm Re}
\lambda^{}_{K^{*+}K^-} - x {\rm Im}
\lambda^{}_{K^{*+}K^-}) \Gamma t ] \; , \nonumber \\
\Gamma [\bar{D}^0(t) \rightarrow K^{*-}K^+] & \propto &
|\bar{A}^{}_{K^{*-}K^+}|^2 e^{-\Gamma t} [ 1 + (y {\rm Re}
\bar{\lambda}^{}_{K^{*-}K^+} - x {\rm Im}
\bar{\lambda}^{}_{K^{*-}K^+}) \Gamma t ] \; ;
\end{eqnarray}
and
\begin{eqnarray}
\Gamma [D^0(t) \rightarrow K^{*-}K^+] & \propto &
|\bar{A}^{}_{K^{*-}K^+}|^2 e^{-\Gamma t} [
|\bar{\lambda}^{}_{K^{*-}K^+}|^2 + (y {\rm Re}
\bar{\lambda}^{}_{K^{*-}K^+} + x {\rm Im}
\bar{\lambda}^{}_{K^{*-}K^+}) \Gamma t ] \left |
\frac{q}{p} \right |^2 , \nonumber \\
\Gamma [\bar{D}^0(t) \rightarrow K^{*+}K^-] & \propto &
|A^{}_{K^{*+}K^-}|^2 e^{-\Gamma t} [ |\lambda^{}_{K^{*+}K^-}|^2 +
(y {\rm Re} \lambda^{}_{K^{*+}K^-} + x {\rm Im}
\lambda^{}_{K^{*+}K^-}) \Gamma t ] \left | \frac{p}{q} \right |^2
, ~~~~
\end{eqnarray}
where we have required $t \leq 1/\Gamma$ for the proper time $t$.
Taking account of Eq. (4) and defining the {\it effective}
$D^0$-$\bar{D}^0$ mixing parameters
\begin{eqnarray}
x^\prime_{\pm} & = & x \cos\delta \pm y \sin\delta \; , \nonumber \\
y^\prime_{\pm} & = & y \cos\delta \pm x \sin\delta \; ,
\end{eqnarray}
we simplify Eqs. (5) and (6) to
\begin{eqnarray}
\Gamma [D^0(t) \rightarrow K^{*+}K^-] & \propto &
|A^{}_{K^{*+}K^-}|^2 e^{-\Gamma t} \left [ 1 + \rho \left
|\frac{q}{p} \right | \left (y^\prime_- \cos\phi - x^\prime_+
\sin\phi \right ) \Gamma t \right ] \; , \nonumber \\
\Gamma [\bar{D}^0(t) \rightarrow K^{*-}K^+] & \propto &
|\bar{A}^{}_{K^{*-}K^+}|^2 e^{-\Gamma t} \left [ 1 + \rho \left
|\frac{p}{q} \right | \left (y^\prime_- \cos\phi + x^\prime_+
\sin\phi \right ) \Gamma t \right ] \; ;
\end{eqnarray}
and
\begin{eqnarray}
\Gamma [D^0(t) \rightarrow K^{*-}K^+] & \propto &
|\bar{A}^{}_{K^{*-}K^+}|^2 e^{-\Gamma t} \left [ \rho^2 + \rho
\left |\frac{q}{p} \right | \left (y^\prime_+ \cos\phi -
x^\prime_- \sin\phi \right ) \Gamma t \right ] \; , \nonumber \\
\Gamma [\bar{D}^0(t) \rightarrow K^{*+}K^-] & \propto &
|A^{}_{K^{*+}K^-}|^2 e^{-\Gamma t} \left [ \rho^2 + \rho \left
|\frac{p}{q} \right | \left (y^\prime_+ \cos\phi + x^\prime_-
\sin\phi \right ) \Gamma t \right ] \; .
\end{eqnarray}
Once these four decay rates are measured, it will be possible to
determine $\rho$ and constrain the magnitudes of both
$D^0$-$\bar{D}^0$ mixing and $CP$ violation. Note that the
deviation of $|p/q|$ (or $|q/p$) from unity, which can also be
determined or constrained from other neutral $D$-meson decays,
signifies $CP$ violation in $D^0$-$\bar{D}^0$ mixing. This effect
is conveniently described by a small parameter $\Delta$ up to the
correction of ${\cal O}(\Delta^2)$; i.e., $p/q = 1 + \Delta$ and
$q/p = 1 - \Delta$. Given $\phi \approx 0$ in the standard model,
useful information on $y^\prime_+$ and $y^\prime_-$ is achievable
from the time-dependent measurements of $D^0$ vs $\bar{D}^0
\rightarrow K^{*\pm}K^\mp$ transitions. A clear difference between
$y^\prime_+$ and $y^\prime_-$ will imply that both $x$ and
$\delta$ are not very small. These points have also been observed
in Ref. \cite{Grossman}.

We remark that the $K^{*\pm}K^\mp$ events of neutral $D$-meson
decays are important, since they can be complementary to the
$K^\pm\pi^\mp$ and $K^+K^-$ (or $\pi^+\pi^-$) events for the
experimental searches for both $D^0$-$\bar{D}^0$ mixing and $CP$
violation. A similar idea, which makes use of the $D^{*\pm}D^\mp$
events of neutral $B$-meson decays to extract the $CP$-violating
phase $\beta$ and test the factorization hypothesis \cite{Xing98},
has actually been adopted by the Belle \cite{BE2} and BaBar
\cite{BB2} Collaborations in their experiments at the KEK and SLAC
$B$ factories.

\vspace{0.5cm}

\framebox{\large\bf 3} ~ Now we turn to the possibility of
measuring coherent $(D^0\bar{D}^0)^{}_C \rightarrow
(K^{*\pm}K^{\mp})(K^{*\pm}K^{\mp})$ decays on the $\psi (3770)$
resonance with $C=-1$ and (or) on the $\psi (4140)$ resonance with
$C=+1$, where $C$ denotes the charge-conjugation parity of the
$D^0$ and $\bar{D}^0$ pair. Both time-dependent and
time-integrated rates of a general $(D^0\bar{D}^0)^{}_C
\rightarrow f^{}_1 f^{}_2$ decay mode, together with their
approximate expressions up to the accuracy of ${\cal O}(x^2)$ and
${\cal O}(y^2)$, have been formulated in Ref. \cite{Xing97}
without special assumptions. Here we focus our interest on the
time-independent measurements of those
$(K^{*\pm}K^{\mp})(K^{*\pm}K^{\mp})$ events from coherent
$(D^0\bar{D}^0)^{}_C$ decays at a high-luminosity $\tau$-charm
factory (e.g., BEPC-II \cite{BEPC}). Let us define $\Gamma^{++}_C
\equiv \Gamma (K^{*+}K^-, K^{*+}K^-)^{}_C$, $\Gamma^{+-}_C \equiv
\Gamma (K^{*+}K^-, K^{*-}K^+)^{}_C$, $\Gamma^{-+}_C \equiv \Gamma
(K^{*-}K^+, K^{*+}K^-)^{}_C$ and $\Gamma^{--}_C \equiv \Gamma
(K^{*-}K^+, K^{*-}K^+)^{}_C$ for four joint decay rates. With the
help of Ref. \cite{Xing97}
\footnote{Note again that $y = -y^{}_D$, where $y^{}_D$ is the
$D^0$-$\bar{D}^0$ mixing parameter defined in Ref.
\cite{Xing97}.},
we explicitly have
\begin{eqnarray}
\Gamma^{++}_C & \propto & 2|A^{}_{K^{*+}K^-}|^4 \left |
\frac{p}{q} \right |^2 \left \{ \left (2 + C \right ) r \left | 1
+ C\lambda^2_{K^{*+}K^-} \right |^2 + \left (1 + C
\right )^2 \left [ |\lambda^{}_{K^{*+}K^-}|^2 \right . \right . \nonumber \\
&& \left . \left . + y \left (1 + |\lambda^{}_{K^{*+}K^-}|^2
\right ) {\rm Re} \lambda^{}_{K^{*+}K^-} + x \left (1 -
|\lambda^{}_{K^{*+}K^-}|^2 \right ) {\rm Im}
\lambda^{}_{K^{*+}K^-} \right ] \right \} \; ,
\nonumber \\
\Gamma^{--}_C & \propto & 2|\bar{A}^{}_{K^{*-}K^+}|^4 \left |
\frac{q}{p} \right |^2 \left \{ \left (2 + C \right ) r \left | 1
+ C\bar{\lambda}^2_{K^{*-}K^+} \right |^2 + \left (1 + C \right
)^2 \left [ |\bar{\lambda}^{}_{K^{*-}K^+}|^2 \right . \right . \nonumber \\
&& \left . \left . + y \left (1 + |\bar{\lambda}^{}_{K^{*-}K^+}|^2
\right ) {\rm Re} \bar{\lambda}^{}_{K^{*-}K^+} + x \left (1 -
|\bar{\lambda}^{}_{K^{*-}K^+}|^2 \right ) {\rm Im}
\bar{\lambda}^{}_{K^{*-}K^+} \right ] \right \} \; ; ~~~~
\end{eqnarray}
and $\Gamma^{-+}_C = \Gamma^{+-}_C$ with
\begin{eqnarray}
\Gamma^{+-}_C & \propto & 2|A^{}_{K^{*+}K^-}|^4 \left \{ \left (2
+ C \right ) r \left | \lambda^{}_{K^{*+}K^-} +
C\bar{\lambda}^{}_{K^{*-}K^+} \right |^2 + \left | 1 + C
\lambda^{}_{K^{*+}K^-}
\bar{\lambda}^{}_{K^{*-}K^+} \right |^2 \right . \nonumber \\
&& \left . + \left (1 + C \right ) y \left [ \left ( 1 +
|\lambda^{}_{K^{*+}K^-}|^2 \right ) {\rm
Re}\bar{\lambda}^{}_{K^{*-}K^+} + \left (1 +
|\bar{\lambda}^{}_{K^{*-}K^+}|^2 \right ) {\rm Re}
\lambda^{}_{K^{*+}K^-} \right ] \right . \nonumber \\
&& \left . - \left (1 + C \right ) x \left [ \left ( 1 -
|\lambda^{}_{K^{*+}K^-}|^2 \right ) {\rm
Im}\bar{\lambda}^{}_{K^{*-}K^+} + \left (1 -
|\bar{\lambda}^{}_{K^{*-}K^+}|^2 \right ) {\rm Im}
\lambda^{}_{K^{*+}K^-} \right ] \right \} \; ,
\end{eqnarray}
where $r \equiv (x^2 + y^2)/2$ is essentially the ratio of
wrong-sign to right-sign events of semileptonic $D^0$ and
$\bar{D}^0$ decays \cite{Xing97,Asner}. When Eq. (4) is taken into
account, Eqs. (10) and (11) can be simplified to
\begin{eqnarray}
\Gamma^{++}_C & \propto & 2|A^{}_{K^{*+}K^-}|^4 \left \{ \left (2
+ C \right ) r \left [ \left | \frac{p}{q} \right |^2 + 2C\rho^2
\cos \left (\delta + \phi \right ) + \rho^4
\left | \frac{q}{p} \right |^2 \right ] \right . \nonumber \\
&& \left . + \left (1 + C \right )^2 \rho \left [ \rho + \left |
\frac{p}{q} \right | \left (y^\prime_+ \cos\phi + x^\prime_-
\sin\phi \right ) + \rho^2 \left | \frac{q}{p} \right | \left
(y^\prime_- \cos\phi - x^\prime_+ \sin\phi \right ) \right ]
\right \} \; , \nonumber \\
\Gamma^{--}_C & \propto & 2|\bar{A}^{}_{K^{*-}K^+}|^4 \left \{
\left (2 + C \right ) r \left [ \left | \frac{q}{p} \right |^2 +
2C\rho^2 \cos \left (\delta - \phi \right
) + \rho^4 \left | \frac{p}{q} \right |^2 \right ] \right . \nonumber \\
&& \left . + \left (1 + C \right )^2 \rho \left [ \rho + \left |
\frac{q}{p} \right | \left (y^\prime_+ \cos\phi - x^\prime_-
\sin\phi \right ) + \rho^2 \left | \frac{p}{q} \right | \left
(y^\prime_- \cos\phi + x^\prime_+ \sin\phi \right ) \right ]
\right \} \; ; ~~~~
\end{eqnarray}
and
\begin{eqnarray}
\Gamma^{+-}_C & \propto & 2 |A^{}_{K^{*+}K^-}|^4 \left \{ \left (2
+ C \right ) r \rho^2 \left [ \left | \frac{p}{q} \right |^2 + 2C
\cos \left (2 \phi \right ) + \left | \frac{q}{p} \right |^2
\right ] + \left [1 + 2C \rho^2 \cos \left (2\delta \right ) +
\rho^4 \right ] \right . \nonumber \\
&& + \left (1 + C \right ) \rho \left | \frac{p}{q} \right | \left
[ \left (y^\prime_- \cos\phi + x^\prime_+ \sin\phi \right ) +
\rho^2 \left (y^\prime_+ \cos\phi + x^\prime_- \sin\phi \right )
\right ] \nonumber \\
&& \left . + \left (1 + C \right ) \rho \left | \frac{q}{p} \right
| \left [ \left (y^\prime_- \cos\phi - x^\prime_+ \sin\phi \right
) + \rho^2 \left (y^\prime_+ \cos\phi - x^\prime_- \sin\phi \right
) \right ] \right \} \; .
\end{eqnarray}
Note that the terms proportional to $r$ in $\Gamma^{\pm\pm}_C$ are
only important when $C=-1$ is taken. As $K^{*+}K^-$ and
$K^{*-}K^+$ are not the $CP$ eigenstates, both $\rho \neq 1$ and
$\delta \neq 0$ are expected to hold. It is therefore reasonable
to neglect the term proportional to $r$ in Eq. (13) even for the
$C=-1$ case. We stress that these formulas will be very useful to
analyze the experimental data on coherent $(D^0\bar{D}^0)^{}_C
\rightarrow (K^{*\pm}K^{\mp})(K^{*\pm}K^{\mp})$ decays at a
$\tau$-charm factory.

For $C=-1$ on the $\psi (3770)$ resonance, we obtain
\begin{eqnarray}
\frac{\Gamma^{++}_-}{\Gamma^{+-}_-} & \approx & r \frac{1 -
2\rho^2 \cos \left (\delta + \phi \right ) + \rho^4}{1 - 2 \rho^2
\cos \left (2\delta \right ) + \rho^4} \; ,
\nonumber \\
\frac{\Gamma^{--}_-}{\Gamma^{+-}_-} & \approx & r \frac{1 -
2\rho^2 \cos \left (\delta - \phi \right ) + \rho^4}{1 - 2 \rho^2
\cos \left (2\delta \right ) + \rho^4} \; ,
\end{eqnarray}
where we have used the approximation $|p/q| \approx |q/p| \approx
1$ and neglected the term proportional to $r$ in $\Gamma^{+-}_-$.
One can clearly see that these two ratios signify
$D^0$-$\bar{D}^0$ mixing (i.e., $r \neq 0$). The difference
between $\Gamma^{++}_-/\Gamma^{+-}_-$ and
$\Gamma^{--}_-/\Gamma^{+-}_-$ measures the $CP$-violating effect
in $D^0$-$\bar{D}^0$ mixing ($\Delta \neq 0$) and that from the
interference between decay and mixing ($\phi \neq 0$):
\begin{equation}
\frac{\Gamma^{++}_- - \Gamma^{--}_-}{\Gamma^{+-}_-} \; \approx \;
\frac{\displaystyle 4 r \left [ \left (1 - \rho^4 \right ) \Delta
+ \rho^2 \sin\delta \sin\phi \right ]}{1 - 2 \rho^2 \cos \left
(2\delta \right ) + \rho^4} \; ,
\end{equation}
where the notations $|p/q| = 1 + \Delta$ and $|q/p| = 1 - \Delta$
have been taken into account. The smallness of $r$ (i.e., $r \sim
10^{-4}$), however, might more or less obstruct the observation of
$\Gamma^{++}_-/\Gamma^{+-}_-$ and $\Gamma^{--}_-/\Gamma^{+-}_-$ at
present. But we hope that the high-luminosity $\tau$-charm factory
may finally realize the desired measurements in the near future.

For $C=+1$ on the $\psi (4140)$ resonance, one may simply neglect
the terms proportional to $r$ in Eqs. (12) and (13). Up to small
corrections of ${\cal O}(x^\prime_\pm)$ and ${\cal
O}(y^\prime_\pm)$, the relationship
\begin{equation}
\frac{\Gamma^{++}_+}{\Gamma^{+-}_+} \; \approx \;
\frac{\Gamma^{--}_+}{\Gamma^{+-}_+} \; \approx \; \frac{4\rho^2}{1
+ 2\rho^2 \cos \left (2\delta \right ) + \rho^4} \;
\end{equation}
holds approximately. Once the ratios $\Gamma^{++}_+/\Gamma^{+-}_+$
and $\Gamma^{--}_+/\Gamma^{+-}_+$ are measured, they will impose a
strong constraint on $\rho$ and $\delta$.  On the other hand, the
difference between $\Gamma^{++}_+/\Gamma^{+-}_+$ and
$\Gamma^{--}_+/\Gamma^{+-}_+$ is a clear signal of $CP$ violation:
\begin{equation}
\frac{\Gamma^{++}_+ - \Gamma^{--}_+}{\Gamma^{+-}_+} \; \approx \;
\frac{\displaystyle 8\rho \left [\Delta \cos\phi \left (y^\prime_+
- \rho^2 y^\prime_- \right ) + \sin\phi \left (y^\prime_- - \rho^2
x^\prime_+ \right ) \right ]}{1 + 2\rho^2 \cos \left (2\delta
\right ) + \rho^4} \; .
\end{equation}
Comparing between $(\Gamma^{++}_- - \Gamma^{--}_-)/\Gamma^{+-}_-$
and $(\Gamma^{++}_+ - \Gamma^{--}_+)/\Gamma^{+-}_+$, we find that
the latter is less suppressed by the smallness of $x$ and $y$.
Hence it seems more promising to measure $CP$ violation in the
decays of correlated $D^0$ and $\bar{D}^0$ mesons into
$(K^{*\pm}K^\mp)(K^{*\pm}K^\mp)$ states on the $\psi (4140)$
resonance.

\vspace{0.5cm}

\framebox{\large\bf 4} ~ Finally let us make some comments on the
final-state interactions in $D^0\rightarrow K^{*\pm}K^\mp$
transitions. A model-independent approach is to do the isospin
analysis of $D^0\rightarrow K^{*+}K^-$, $D^0\rightarrow
\bar{K}^{*0}K^0$ and $D^+\rightarrow K^{*+}\bar{K}^0$ decays or
$D^0\rightarrow K^{*-}K^+$, $D^0\rightarrow K^{*0}\bar{K}^0$ and
$D^+\rightarrow \bar{K}^{*0}K^+$ decays, in which each final state
contains $I=0$ and (or) $I=1$ isospin configurations. For
simplicity, we denote the amplitudes of $D^0\rightarrow
K^{*+}K^-$, $D^0\rightarrow \bar{K}^{*0}K^0$ and $D^+\rightarrow
K^{*+}\bar{K}^0$ as $A^{}_{K^{*+}K^-}$, $A^{}_{\bar{K}^{*0}K^0}$
and $A^{}_{K^{*+}\bar{K}^0}$, respectively. They can be expressed
in terms of two independent isospin amplitudes $A^{}_0$ and
$A^{}_1$ as follows \cite{Xing97}:
\begin{eqnarray}
A^{}_{K^{*+}K^-} & = & \frac{1}{2} \left ( A^{}_1 + A^{}_0 \right
) \; , \nonumber \\
A^{}_{\bar{K}^{*0}K^0} & = & \frac{1}{2} \left ( A^{}_1 - A^{}_0
\right ) \; , \nonumber \\
A^{}_{K^{*+}\bar{K}^0} & = & A^{}_1 \; .
\end{eqnarray}
The branching ratios of these three decays are $B^{}_{K^{*+}K^-}
\propto |A^{}_{K^{*+}K^-}|^2 \tau^{}_0$, $B^{}_{\bar{K}^{*0}K^0}
\propto |A^{}_{\bar{K}^{*0}K^0}|^2 \tau^{}_0$ and
$B^{}_{K^{*+}\bar{K}^0} \propto |A^{}_{K^{*+}\bar{K}^0}|^2
\tau^{}_+$, where $\tau^{}_{0} = (410.1 \pm 1.5) \times
10^{-15}~{\rm s}$ and $\tau^{}_+ = (1040 \pm 7) \times
10^{-15}~{\rm s}$ are the lifetimes of $D^0$ and $D^+$ mesons
\cite{PDG06}, respectively. Defining $A^{}_0/A^{}_1 = z
e^{i\varphi}$, we find
\begin{eqnarray}
\left | \frac{A^{}_{K^{*+}K^-}}{A^{}_{K^{*+}\bar{K}^0}} \right |^2 &
= & \frac{\tau^{}_+}{\tau^{}_0} \cdot \frac{B^{}_{K^{*+}
K^-}}{B^{}_{K^{*+}\bar{K}^0}} \; = \; \frac{1}{4} \left |1 + z
e^{i\varphi} \right |^2 \; ,
\nonumber \\
\left | \frac{A^{}_{\bar{K}^{*0}K^0}}{A^{}_{K^{*+}\bar{K}^0}}
\right|^2 & = & \frac{\tau^{}_+}{\tau^{}_0} \cdot
\frac{B^{}_{\bar{K}^{*0}K^0}}{B^{}_{K^{*+}\bar{K}^0}} \; = \;
\frac{1}{4} \left |1 - z e^{i\varphi} \right |^2 \; .
\end{eqnarray}
Then the isospin parameters $z$ and $\varphi$ can be determined:
\begin{eqnarray}
z & = & \left [2 \frac{\tau^{}_+}{\tau^{}_0} \cdot
\frac{B^{}_{K^{*+}K^-} +
B^{}_{\bar{K}^{*0}K^0}}{B^{}_{K^{*+}\bar{K}^0}} - 1 \right ]^{1/2}
\; , \nonumber \\
\varphi & = & \arccos \left( \frac{\tau^{}_+}{\tau^{}_0} \cdot
\frac{B^{}_{K^{*+}K^-} - B^{}_{\bar{K}^{*0}K^0}}{z
B^{}_{K^{*+}\bar{K}^0}} \right) \; .
\end{eqnarray}
Of course, $\varphi \neq 0$ implies the existence of final-state
interactions. One may follow a similar procedure to do the isospin
analysis of $D^0\rightarrow K^{*-}K^+$, $D^0\rightarrow
K^{*0}\bar{K}^0$ and $D^+\rightarrow \bar{K}^{*0}K^+$ decays. The
amplitudes of these three transitions are essentially identical to
those of $\bar{D}^0\rightarrow K^{*+}K^-$, $\bar{D}^0\rightarrow
\bar{K}^{*0}K^0$ and $D^-\rightarrow K^{*0}K^-$ transitions, since
their tree-level quark diagrams are $CP$-conserving and the
penguin diagrams are negligibly small in the standard model. The
corresponding isospin parameters $\bar{z}$ and $\bar{\varphi}$ can
be extracted from the branching ratios $B^{}_{K^{*-}K^+}$,
$B^{}_{K^{*0}\bar{K}^0}$ and $B^{}_{\bar{K}^{*0}K^+}$ in the
$CP$-conserving case. It is in general difficult to link the
isospin phase differences $\varphi$ and $\bar{\varphi}$ to the
strong phase difference $\delta$ defined in Eq. (2), unless some
assumptions are made in a specific model of hadronic matrix
elements. Nevertheless, it is reasonable to argue that significant
$\varphi$ and $\bar{\varphi}$ must hint at significant $\delta$
for $K^{*\pm}K^\mp$ events.

For the purpose of illustration, we do a numerical analysis of the
isospin parameters by using the present experimental data
\cite{PDG06},
\begin{eqnarray}
B^{}_{K^{*+} \bar{K}^0} & = & (3.2 \pm 1.4) \times 10^{-2}
\; , \nonumber \\
B^{}_{K^{*+} K^-} & = & (3.7 \pm 0.8) \times 10^{-3} \; ,
\nonumber \\
B^{}_{\bar{K}^{*0} K^0} & < & 1.6 \times 10^{-3} \; ;
\end{eqnarray}
and
\begin{eqnarray}
B^{}_{\bar{K}^{*0} K^+} & = & (3.02 \pm 0.35) \times 10^{-3}
\; , \nonumber \\
B^{}_{K^{*-} K^+} & = & (2.0 \pm 1.1) \times 10^{-3} \; ,
\nonumber \\
B^{}_{K^{*0} \bar{K}^0} & < & 8 \times 10^{-4} \; .
\end{eqnarray}
Since the magnitudes of $B^{}_{\bar{K}^{*0} K^0}$ and
$B^{}_{K^{*0} \bar{K}^0}$ have not been fixed, our analysis can
only provide some limited information on $(z, \varphi)$ and
$(\bar{z}, \bar{\varphi})$. The numerical results are shown in
Fig. 1 and Fig. 2. Some comments are in order:
\begin{enumerate}
\item It is straightforward to see that the possibility of
$\varphi =0$ and (or) $\bar{\varphi} =0$ is almost excluded by
current experimental data. The most favorable values of $\varphi$
and $\bar{\varphi}$ are around $50^\circ$, implying the presence
of significant final-state interactions. Indeed, $\varphi$ can be
as large as $65^\circ$, and $\bar{\varphi}$ can be even larger
than $80^\circ$. The strong phase difference $\delta$ is therefore
expected to be significant in $D^0$ vs $\bar{D}^0\rightarrow
K^{*\pm}K^\mp$ transitions.

\item The constraints on
$B^{}_{\bar{K}^{*0} K^0}/B^{}_{K^{*+} \bar{K}^0}$ and
$B^{}_{K^{*0} \bar{K}^{0}}/B^{}_{\bar{K}^{*0} K^+}$ allow us to
extract the lower and (or) upper bounds of $B^{}_{\bar{K}^{*0}
K^0}$ and $B^{}_{K^{*0} \bar{K}^{0}}$. We find $3.2 \times 10^{-4}
\leq B^{}_{\bar{K}^{*0} K^0} \leq 1.6 \times 10^{-3}$ and
$B^{}_{K^{*0} \bar{K}^{0}} < 8 \times 10^{-4}$. The former is
interesting and can be tested in the upcoming experiments, but the
latter is trivial. More accurate data will reduce the
uncertainties in our isospin analysis.

\item The allowed ranges of $z$ and $\bar{z}$ do not have much overlap.
In particular,
$\bar{z} > 0.8 >z$ is roughly true. This observation implies that
$D^0 \rightarrow K^{*\pm}K^\mp$ and $\bar{D}^0 \rightarrow
K^{*\pm}K^\mp$ decays might involve quite different final-state
interactions, from which significant $\delta$ is naturally
anticipated.
\end{enumerate}
It is worth mentioning that an isospin analysis of $D^0\rightarrow
K^+K^-$, $D^0\rightarrow K^0\bar{K}^0$ and $D^+\rightarrow
K^+\bar{K}^0$ decays \cite{Xing97}, whose branching ratios have
all been measured, also indicates the existence of strong
final-state interactions. As $K^+K^-$ is a $CP$ eigenstate,
however, the ratio of $\langle K^+K^-|{\cal H}|\bar{D}^0\rangle$
to $\langle K^+K^-|{\cal H}|D^0\rangle$ does not involve a
significant strong phase difference in the absence of direct $CP$
violation \cite{Nir}.

\vspace{0.5cm}

\framebox{\large\bf 5} ~ In summary, we have investigated
$D^0$-$\bar{D}^0$ mixing and $CP$ violation in $D^0 \rightarrow
K^{*\pm} K^\mp$ decays and their $CP$-conjugate processes, whose
final states may have a significant strong phase difference. We
have shown that both the $D^0$-$\bar{D}^0$ mixing parameters ($x$
and $y$) and the strong phase difference ($\delta$) can be
determined or constrained from the time-dependent measurements of
$D^0$ vs $\bar{D}^0 \rightarrow K^{*\pm} K^\mp$ decays. For a
high-luminosity $\tau$-charm factory running on the $\psi (3770)$
and $\psi (4140)$ resonances, we find that it is even possible to
determine or constrain $x$, $y$ and $\delta$ from the
time-independent measurements of coherent $(D^0\bar{D}^0)
\rightarrow (K^{*\pm} K^\mp)(K^{*\pm} K^\mp)$ events. If the
$CP$-violating phase of $D^0$-$\bar{D}^0$ mixing is significant in
a scenario beyond the standard model, it can also be extracted
from the decay modes under discussion.

We strongly recommend the experimentalists to pay some special
attention to the $K^{*\pm}K^\mp$ events of neutral $D$-meson
decays, because they are complementary to the $K^\pm\pi^\mp$ and
$K^+K^-$ (or $\pi^+\pi^-$) events for the study of both
$D^0$-$\bar{D}^0$ mixing and $CP$ violation. We expect that these
interesting channels and possible new physics in them can well be
explored at BEPC-II and other charm-physics experiments in the
near future.

\vspace{0.5cm}

One of us (Z.Z.X.) would like to thank Phil Chan and the National
University of Singapore for warm hospitality, where this paper was
written. He is also grateful to H.B. Li for some interesting
discussions about the BaBar and Belle results. This work was
supported in part by the National Natural Science Foundation of
China.

\newpage

\begin{figure}[t]
\vspace{-9cm}
\epsfig{file=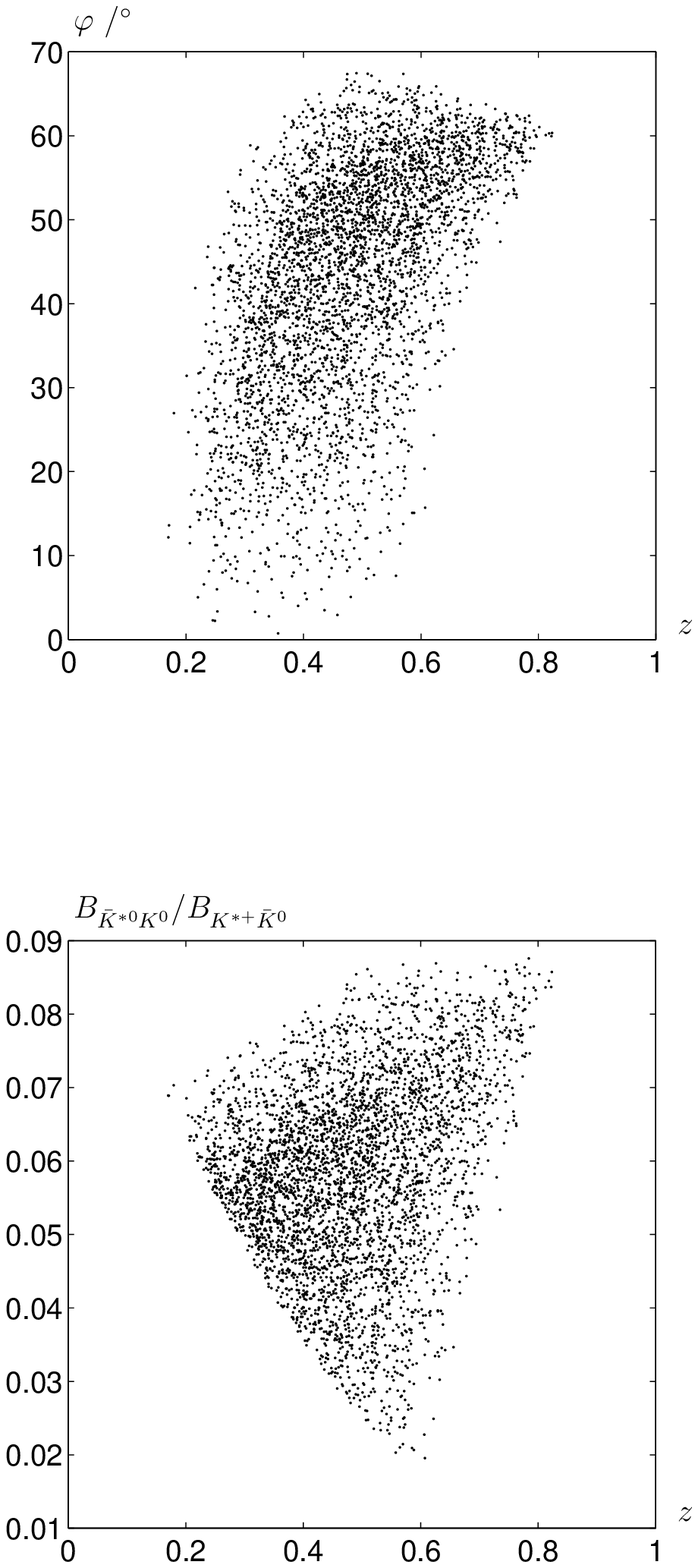,bbllx=0.5cm,bblly=14cm,bburx=12.5cm,bbury=34cm,%
width=10cm,height=16cm,angle=0,clip=0} \vspace{7.5cm}
\caption{Numerical illustration of the allowed ranges of $z$,
$\varphi$ and $B^{}_{\bar{K}^{*0} K^0}/B^{}_{K^{*+} \bar{K}^0}$.}
\end{figure}

\begin{figure}[t]
\vspace{-9cm}
\epsfig{file=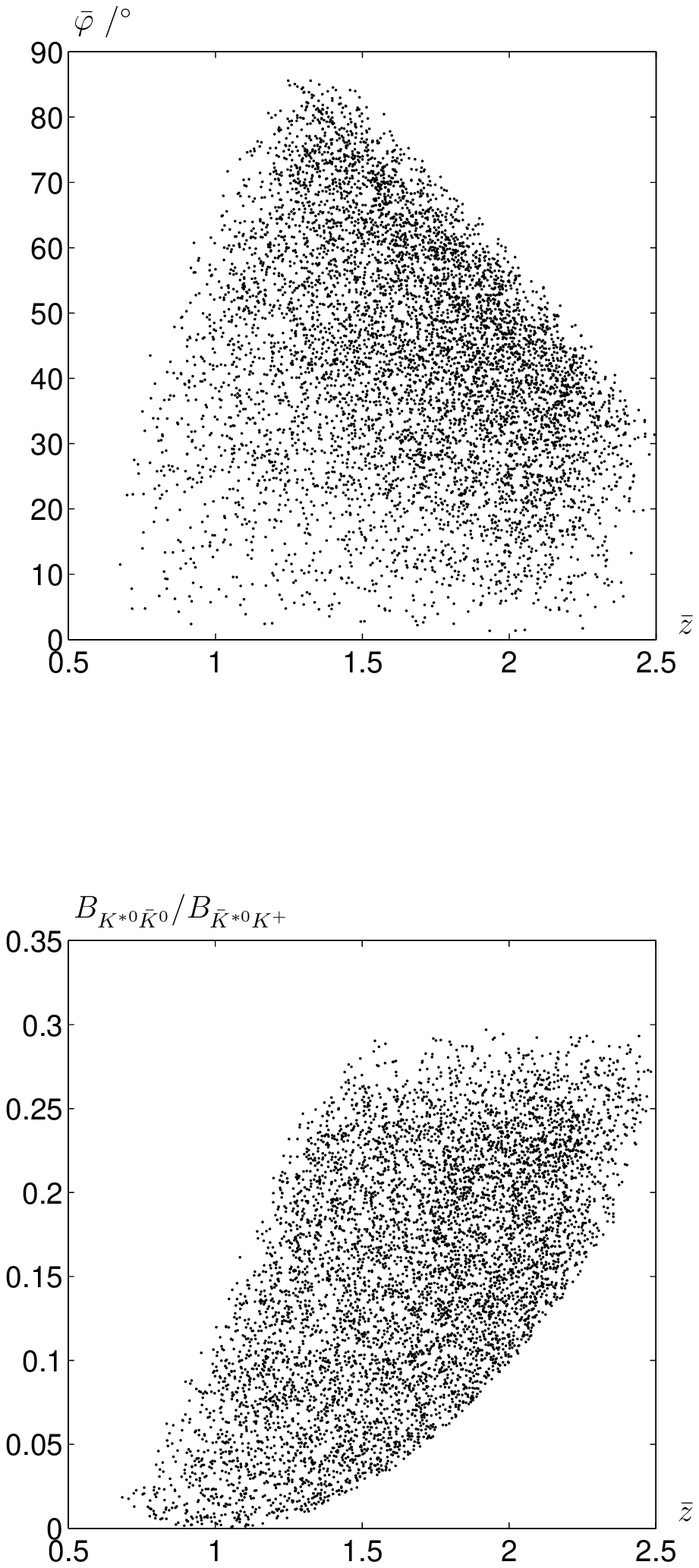,bbllx=0.5cm,bblly=14cm,bburx=12.5cm,bbury=34cm,%
width=10cm,height=16cm,angle=0,clip=0} \vspace{7.5cm}
\caption{Numerical illustration of the allowed ranges of
$\bar{z}$, $\bar{\varphi}$ and $B^{}_{K^{*0}
\bar{K}^{0}}/B^{}_{\bar{K}^{*0} K^+}$.}
\end{figure}


\vspace{0.5cm}

\begin{thebibliography}{99}

\bibitem{BB} BaBar Collaboration, B. Aubert {\it et al.},
hep-ex/0703020.

\bibitem{BE} Belle Collaboration, K. Abe {\it et al.},
hep-ex/0703036.

\bibitem{Nir} Y. Nir, hep-ph/0703235.

\bibitem{Italy} M. Ciuchini, E. Franco, D. Guadagnoli,
V. Lubicz, M. Pierini, V. Porretti, and L. Silverstrini,
hep-ph/0703204.

\bibitem{Buras} M. Blanke, A.J. Buras, S. Recksiegel, C. Tarantino,
and S. Uhlig, hep-ph/0703254.

\bibitem{He} X.G. He and G. Valencia, hep-ph/0703270.

\bibitem{Li} X.D. Cheng, K.L. He, H.B. Li, Y.F. Wang, and M.Z. Yang,
arXiv:0704.0120.

\bibitem{Chen} C.H. Chen, C.Q. Geng, and T.C. Yuan,
arXiv:0704.0601.

\bibitem{Ball} P. Ball, arXiv:0704.0786.

\bibitem{LD} For a recent review with extensive references, see:
A.A. Petrov, Int. J. Mod. Phys. A {\bf 21}, 5686 (2006).

\bibitem{Petrov} A.F. Falk, Y. Grossman, Z. Ligeti, and A.A.
Petrov, Phys. Rev. D {\bf 65}, 054034 (2002).

\bibitem{Burdman} G. Burdman and I. Shipsey, Ann. Rev. Nucl. Part.
Sci. {\bf 53}, 431 (2003).

\bibitem{BEPC} Y.F. Wang {\it et al.}, in {\it Proceedings
of the International Workshop on Tau-Charm Physics}, Beijing,
2006, edited by Y.F. Wang and H.B. Li, Int. J. Mod. Phys. A {\bf
21}, 5371 (2006).

\bibitem{Xing96} Z.Z. Xing, Phys. Lett. B {\bf 372}, 317 (1996).

\bibitem{Grossman} Y. Grossman, A.L. Kagan, and Y. Nir,
Phys. Rev. D {\bf 75}, 036008 (2007); and references therein.

\bibitem{Xing97} Z.Z. Xing, Phys. Rev. D {\bf 55}, 196 (1997).

\bibitem{Asner} I.I. Bigi and A.I. Sanda, Phys. Lett. B {\bf 171},
320 (1986); D. Du, Phys. Rev. D {\bf 34}, 3428 (1986); Z.Z. Xing,
Phys. Lett. B {\bf 379}, 257 (1996); D.M. Asner and W.M. Sun,
Phys. Rev. D {\bf 73}, 034024 (2006).

\bibitem{Xing98} Z.Z. Xing, Phys. Lett. B {\bf 443}, 365 (1998);
Phys. Rev. D {\bf 61}, 014010 (2000); X.Y. Pham and Z.Z. Xing,
Phys. Lett. B {\bf 458}, 375 (1999).

\bibitem{BE2} Belle Collaboration, K. Abe {\it et al.}, Phys. Rev.
Lett. {\bf 89}, 122001 (2002); T. Aushev {\it et al.}, Phys. Rev.
Lett. {\bf 93}, 201802 (2004).

\bibitem{BB2} BaBar Collaboration, B. Aubert {\it et al.},
Phys. Rev. Lett. {\bf 95}, 131802 (2005); Phys. Rev. D {\bf 73},
112004 (2006).

\bibitem{PDG06} Particle Data Group, W.M. Yao {\it et al.},
J. Phys. G {\bf 33}, 1 (2006).

\end{thebibliography}
\end{document}